\documentclass[floatfix,aps,showpacs,preprint,nofootinbib,preprintnumbers]
{revtex4}
\usepackage{graphicx}
\begin{document}

\title{Clustering of Primordial Black Holes: Basic Results}

\preprint{astro-ph/0509141}

\author{James R. Chisholm}
\affiliation{Institute for Fundamental Theory, University of Florida, Gainesville, FL 32611-8440,\\
Particle Astrophysics Center, Fermi National Accelerator Laboratory, 
Batavia, Illinois 60510-0500, and \\ Enrico Fermi Institute, University of Chicago, Illinois 60637}
\email{chisholm@phys.ufl.edu}
\date{\today}

\begin{abstract}
We investigate the spatial clustering properties of primordial black holes (PBHs).  
With minimal assumptions, we show that PBHs created in the radiation era are highly clustered.  
Using the peaks theory model of bias, we compute the PBH two-point correlation function and power spectrum.  
For creation from an initially adiabatic power spectrum of perturbations, the PBH power spectrum contains both isocurvature and adiabatic components.  
The absence of observed isocurvature fluctuations today constrains the mass range in which PBHs may serve as dark matter.  
We briefly discuss other consequences of PBH clustering.
\end{abstract}


\maketitle

\section{Introduction}

Primordial black holes (PBHs) are a unique probe of cosmology, general relativity, and quantum gravity.  
Formed by high concentrations of energy density in the early universe, PBHs are distinguished from other 
(astrophysical) black holes by not being formed through stellar collapse.  
In this paper we concentrate on PBHs formed from the direct gravitational collapse of density perturbations 
that are of order unity on the scale of the cosmological horizon \cite{zeldovich,hawking1} upon horizon entry,
though there are other
mechanisms for their creation, {\it e.g.} collapse of cosmic strings \cite{polnarev} or domain walls \cite{rubin},
or from bubble collisions \cite{HMS} in the early universe.
 
Measurements of the cosmic microwave background (CMB) anisotropy \cite{WMAP1} imply that density
perturbations at the time of decoupling are much smaller ($\delta_H \approx 10^{-5}$).  As such, 
PBH formation will be cosmologically negligible during and beyond this era.  Less constrained are
the conditions in the early universe before decoupling, and we cannot preclude the existence of much
larger density contrasts which could have formed PBHs.

The theory of inflation \cite{liddlelyth} has been successful in describing both the large-scale homogeneity of the
universe and the formation of small-scale structure through the creation of a spectrum of cosmological 
perturbations.  It predicts an era of accelerated expansion dominated by
the energy of a slowly rolling scalar field, ending in a period of reheating where the energy density is transferred
into (more or less) the particles we observe today and the radiation dominated epoch begins.  The period of reheating
is important for PBH production in two ways.  First, it is the highest energy scale at which one would expect PBH to 
take place.  Gravitational collapse is inhibited by the accelerated expansion, and the number density of any PBHs that do 
form would be drastically diluted.  Second, several models of inflation exhibit an increase in the amplitude
of perturbations at the end of inflation (at the epoch of reheating), which increases the probability of PBH formation.

One topic of interest is the feasibility of PBHs as dark matter (DM) \cite{chapline}.
PBHs appear to be an {\it a priori} good CDM candidate.  Formed purely by gravity, they require no 
special extensions to the Standard Model of Particle Physics (such as supersymmetry), and are predicted
on quite generic grounds to form in the early universe \cite{hawking1}.  While the smaller masses of 
PBHs (compared to astrophysical black holes) mean that Hawking radiation is non-negligible, PBHs that
are still in the present universe are still ``dark'' like other BHs.

Because of this, there have been a number of studies of PBHs as CDM in the literature.  We can
split them roughly into three categories:

{\bf QCD PBHs:} These are PBHs formed during the QCD phase transition,
being a fraction of a solar mass \cite{jedamzik, ichiki}.  
This was initially attractive as evidence from microlensing
events suggested a population of MACHOs in just this mass range.  However, in order to produce
the correct $\Omega_m$, one needs to invoke a ``blue'' spectrum ($n > 1$) of perturbations, which
is highly disfavored by CMB observations.  Further, the evidence that these MACHOS compromise
a substantial fraction of DM halos is lessening \cite{GreJed}.

{\bf ``Spiky'' PBHs:} These are PBHs formed due to the enhancement of power below a certain
scale due to features (such as spikes) in the radiation power spectrum \cite{yokoyama3, ivanov2, ivanov, blais, blais2}.  
Such a ``spiky'' power spectrum (a generalizion of a ``blue'' spectrum, where just the 
power-law slope is changed) can be produced in inflationary models with ``plateaus'' in the
inflationary potential.  PBHs created in this manner can exist over a larger range of masses,
given the increased freedom in choosing an inflationary model.  Included in this class are PBHs
created due to perturbation amplification due to preheating \cite{garcia2, bassett, GreMal, Suyama}.

{\bf Relic PBHs:} These are PBHs of around a Planck mass that exist in some theories
of quantum gravity as the end result of PBH evaporation \cite{macgibbon, barrow, barrau}.  
As all PBHs with initial masses less than $\sim 10^{15}$ g would have evaporated by the
present day, any model that produces a number of light PBHs will leave behind relic PBHs.

The only limits on PBHs with masses above $10^{15}$ g derive from the requirement that they do
not overclose the universe ($\Omega_{PBH} < 1$), so there is a range of PBH masses over which
they may serve as DM.  Knowing the PBH abundance is necessary, but not sufficient, to fully
gauge their feasibility as DM.  Also important are their spatial clustering properties, as that too
is constrained by CMB and large scale structure (LSS) data, though to date discussions of PBH 
clustering have been sparse in the literature.  A recent general review of PBHs can be found in \cite{carrreview}.

\subsection{PBH Clustering}

The first discussions of PBH clustering came soon after their ``discovery''.  A theory
was posited by \citet{meszaros} where galaxy formation proceeds from the fluctuations 
in PBH number density.  The model does not address how the PBHs are created, but assumes
they are around a solar mass and created at or before the QCD phase transition.  It 
claimed that for PBH fluctuations that are uncorrelated on scales greater than the
horizon scale (i.e., Poisson fluctuations only), it would be sufficient to able to 
allow for galaxy formation.  This model was refuted in \cite{carr5} (and later expanded upon
in \cite{carr11}), where it was pointed
out that the PBH creation process cannot create the ``extra'' density fluctuations on
super-horizon scales that was claimed\footnote{Though see \cite{meszaros2} for a refutation of some of
the refutations of \cite{carr5} and \cite{carr11}.}.

\citet{kotok} posit a theory where an initial stage (1st generation) of PBH formation leads to an early stage 
of matter (PBH) domination.  PBH clustering then enhances a second stage (2nd generation) of PBH formation 
due to collapse
in this (pressureless) era; specifically, due to the coagulation of PBHs during matter domination.
Provided this coagulation is not complete, the remainder of the 1st generation PBHs evaporate (thus, reheating
the universe) leaving behind the 2nd generation of PBHs.  They claim that with a ``blue'' spectrum of initial 
perturbations ($n \geq 1.2$), PBHs of the 2nd generation are overproduced with respect to 
observational constraints.

While PBH reheating has been considered \citep{garcia,garcia2}, it can be shown that \citep{lemoine,green2,khlopov}
that the period of PBH domination necessary would lead to the overproduction of (supersymmetric) moduli fields and
gravitinos upon their evaporation that contradict the predictions of big bang nucleosynthesis (BBN).  While
the authors of \cite{kotok} seem to confuse the distinction between radiation perturbations and PBH 
perturbations (see their Equation (11)), we show later that PBH merging could be a natural consequence of clustering.

Assuming PBHs comprise the bulk of the CDM, \citet{afshordi} study how the discreteness of their
population affects the CDM power spectrum.  They note that PBH perturbations
on large scales (super-horizon sized at creation) are a mixture of adiabatic (as with other forms of CDM)
and isocurvature (due to Poisson fluctuations alone).  Using Ly$\alpha$ forest observations, they use this to 
constrain the mass of PBHs to be less than $10^4 M_\odot$.  They are also the first to investigate PBH
cluster dynamics; estimating the lifetime due to ``evaporation'' (different from Hawking evaporation) to
show that PBH clusters with $N \lesssim 3000$ objects will evaporate by the current day.  We expand on this
analysis later.

Results from microlensing experiments indicate a population of 
Massive Compact Halo Objects (MACHOs) in our galaxy.  A possibility that this population
is made up of PBHs of around a half a solar mass, the right mass range for QCD PBHs.
In such a population, gravitational attraction between
PBHs would induce the formation of PBH-PBH binaries.  As such, such objects have been studied
as sources of gravitational waves \cite{nakamura, ioka1, hiscock, ioka2}, though to date no
such signals have been detected \cite{abbott}.

PBHs would be the first gravitationally collapsed objects in the universe.  As clustering
is ubiquitous in other, observed gravitationally collapsed systems (galaxies, clusters of 
galaxies, superclusters, etc), it will be no different for PBHs.  The aim of this work is 
to compute the spatial clustering properties of PBHs, and see what impact that has for 
PBHs in cosmology.  We will be particularly interested in the viability of PBHs as DM.
In Section~\ref{basics} we describe general properties of PBHs we will use throughout the paper.
In Section~\ref{bias} we derive the initial clustering properties of PBHs after their
formation, computing the PBH two-point correlation function and power spectrum.  We conclude in 
Section~\ref{conc} with a discussion of observational constraints and avenues for
further research.

\section{PBH Basics}\label{basics}

A black hole of mass $M$ has a Schwarzschild radius $R_S = 2GM = \frac{2M}{M_P^2}$.  Throughout
we assume that any PBHs have negligible angular momentum and electric charge.

PBHs form from large perturbations in the radiation density field that are able to overcome
the resistance of radiation pressure and collapse directly to black holes.  For a perturbation
of a fixed comoving size, it cannot begin to collapse until it passes within the cosmological
horizon.  The size of a PBH when it forms, therefore, is related to the horizon size when 
the collapsing perturbation enters the horizon\footnote{Which is to be expected, being the
only characteristic length scale involved.}.  In the radiation dominated regime where $a 
\propto t^{1/2}$ and assuming a top-hat window function, the horizon mass is simply
\begin{equation}\label{horizonmass}
M_H(t) = M_P \left(\frac{t}{t_P}\right) = (2\times10^{5} M_\odot) \left(\frac{t}{1 \textrm{s}}\right)
\end{equation}
where $t_P$ is the Planck time.  Assuming radiation domination, we can rewrite this in terms of
temperature as
\begin{equation}
M_H(T) = \left(\frac{3\sqrt{5}}{4 \pi^{3/2} g_*^{1/2}}\right)\left(\frac{T}{M_P}\right)^{-2} M_P \approx 10^{18} \textrm{g} \left(\frac{T}{10^7 \textrm{GeV}}\right)^{-2}
\end{equation}
where $g_*$ is the effective number of relativistic degrees of freedom. 

The Hubble scale is then determined by the Friedmann equation
\begin{equation}
\label{Friedmann}
H^2 = \frac{8\pi G}{3} \rho.
\end{equation}

The fluctuation of the (radiation) density field is defined as
\begin{equation}
\delta = \frac{\rho - \bar \rho}{\bar \rho},
\end{equation}
and is characterized by its variance on a comoving scale $\chi$ at a time $t$ as
\begin{equation}
\label{sigmasquared}
\sigma^2(\chi,t) \equiv \frac{V}{2\pi^2} \int dk k^2 P(k) T(k,t)^2 |W_k(k\chi)|^2 
\end{equation}
where $P(k)$ is the (primordial) power spectrum, $W_k$ is the Fourier transform of the window function, and 
$T(k,t)$ is the transfer function appropriate for the type of perturbation (adiabatic or isocurvature).  We
further assume that the perturbations are gaussian.  It is known that the perturbations cannot be completely gaussian,
as that would predict perturbations with $\delta < -1$, implying negative energy densities.  This non-gaussianity is 
especially important in the production of PBHs \cite{jedamzik}, as they derive from the high end (tail) of the
probability distribution.  Nevertheless, we focus here on the case of underlying gaussian perturbations for 
computational ease.

Consider a perturbation $\delta(r_H)$ smoothed over the comoving Hubble radius $r_H = R_H/a = (aH)^{-1}$.  
As the underlying perturbations
$\delta_k$ are assumed gaussian, the smoothed perturbation will be as well (central limit theorem).
As the perturbation must have enough mass to overcome pressure, there is a threshold value $\delta_c$ 
below which a PBH will not form.  Further, the horizon sized perturbation cannot be larger than unity,
or it will pinch off and form a separate universe \cite{carr8}.  Therefore the range that forms
PBHs is $\delta \in [\delta_c,1]$.
The exact value of $\delta_c$ is not known precisely.  Analytically, $\delta_c = w$, where
$w$ is the equation of state parameter of the background universe defined through $p=w\rho$.
The PBH mass was then estimated to be $M = w^{3/2} M_H$.  Numerical simulations of PBH formation \cite{niemeyer,hawke,musco}
have shown a more complex relation, where
\begin{equation}
M = \kappa M_H (\delta - \delta_c)^\gamma,
\end{equation}
in accordance with other critical phenomena, where $\kappa \approx 3$, $\gamma \approx 0.7$ and
$\delta_c \approx 2 w$.  The values of these parameters vary depending on the shape of the 
initial perturbation (gaussian, polynomial, etc.).  This formally allows for PBHs of an 
arbitrarily small mass compared to the horizon size, though some numerical simulations \cite{hawke} have
showed that there is a minimum value of $\approx 10^{-3.5} M_H$ as $\delta \rightarrow \delta_c$.
Rather than focus on one particular formula, we can encapsulate our uncertainty in the PBH mass-horizon 
mass relation with a parameter $f$:
\begin{equation}\label{pbh_horizon_mass_relation}
M_{PBH} = f(w,\delta_c, t, ...) M_H
\end{equation}


While the study of single PBH creation is numerically tractable, the same is not true for studying the
PBH population as a whole due to their incredible rarity.  As such we resort to analytical estimates.
Given a creation threshold $\delta_c$ and the value of the radiation fluctuation size at horizon crossing
$\sigma_{rad}(r_H)$, the probability of forming a PBH within a given horizon volume 
is simply the probability of
having a perturbation with $\delta_c < \delta < 1$, or
\begin{equation}
\beta = \int_{\delta_c}^{1} \left(2 \pi \sigma_{rad}^2(r_H)\right)^{-1/2} \exp \left(- \frac{\delta^2}{2 \sigma_{rad}^2(r_H)}\right) d\delta
\end{equation}
Introducing $\nu = \delta_c/\sigma_{rad}(r_H)$ (the threshold in ``sigma'' units);
in the limit where $\nu = \delta_c/\sigma(r_H) >> 1$, the upper limit can be taken to infinity,
so that the expression can be written in terms of the complementary error function (erfc) as
\begin{equation}
\label{beta}
\beta = \textrm{erfc} \left(\frac{\nu}{\sqrt{2}}\right) \approx \sqrt{\frac{2}{\pi}}\frac{e^{-\nu^2/2}}{\nu}
\end{equation}

Note that this can be used to determine the initial PBH density\footnote{Sometimes quoted in the literature
instead of $\beta$ is $\alpha = \beta/(1-\beta)$.  In the limit where the PBH mass $M \approx M_H$, the initial $\rho_{PBH}/\rho_{rad} = \alpha$. }
\begin{equation}
\label{omega}
\Omega_{PBH}(\nu,M) = \frac{\rho_{PBH}}{\rho_c} = \beta \left(\frac{M}{M_H}\right) = f \beta \equiv B
\end{equation}
where we have used $n_{PBH} \equiv \beta/V_H$.
 
Without an observed population to compare calculations to, the value of the (physical) PBH number density varies
in the literature.  While \cite{BBKS} did not address PBH formation {\it per se}, knowing that PBHs 
form at 
peaks in the density field implies
\begin{equation}
\label{bbks}
n_{PBH} = \frac{(n+3)^{3/2}}{(2\pi)^{2} 6^{3/2}} \left(\nu^2 - 1 \right) e^{-\nu^2/2} R_H^{-3}
\end{equation}
where $n$ is the index of the power spectrum ($n=1$ for a scale-invariant spectrum).  Whereas \cite{BBKS} use peaks in the density field, \cite{GLMS} uses peaks in the metric 
perturbation to compute a density which is identical to the \cite{BBKS} result, but with $(n+3)$ replaced
with $(n-1)$.  This latter calculation only holds for $n > 1$ \cite{green_private_communication}.
Generically, we can write the initial PBH density as
\begin{equation}
n_{PBH} = \frac{N_*(\nu) e^{-\nu^2/2}}{V_W}
\end{equation}
where $N_*(\nu)$ encapsulates the non-exponential dependence upon $\nu$.  Equivalently, the initial horizon fraction
going into PBHs is
\begin{equation}
\beta = N_*(\nu) e^{-\nu^2/2}.
\end{equation}
The values of $N_*(\nu)$ for the different models is summarized below:
\begin{eqnarray}
\label{nstar}
&& N_*(\nu) = \left\{\begin{array}{cl}
\sqrt{\frac{2}{\pi}}\nu^{-1} & \textrm{erfc approximation}\\
\frac{1}{\sqrt{2\pi}} \left(\frac{n+3}{6}\right)^{3/2} \left(\nu^2-1\right) & \textrm{BBKS}\\
\frac{1}{\sqrt{2\pi}} \left(\frac{n-1}{6}\right)^{3/2} \left(\nu^2-1\right) & \textrm{GLMS}\\
\end{array} \right\}
\end{eqnarray}

\begin{figure}
\includegraphics[width=5.75in]{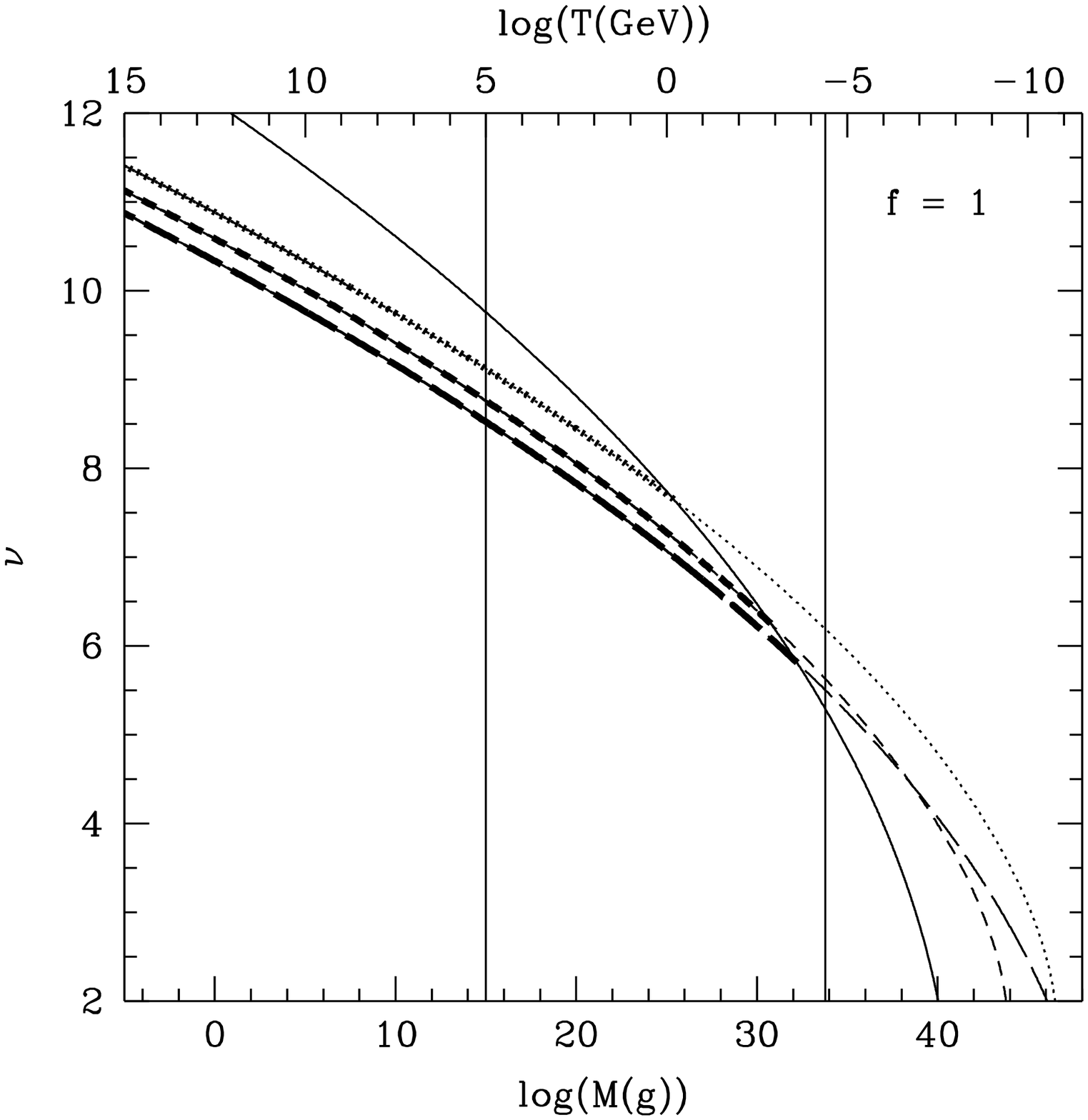}
\caption[Allowed region in $\nu-M_{PBH}$ space: $f=1$]{Allowed region in $\nu - M_{PBH}$ parameter space for PBHs, assuming $f=1$.  Solid curve is the upper limit on $\nu$ due to isocurvature perturbations (from Equation~(\ref{fluclimit})).  Other curves are lower limits on $\nu$ due to number density (Equation~(\ref{nstar})): long dashed line uses the erfc approximation, dotted line uses the BBKS formula with $n=1$, short dashed line uses the GLMS formula with $n=1.5$.  Heavy lines show where PBH dark matter is allowed by the isocurvature constraint.  Shown also is the temperature of the universe $T$ when PBHs form.  The line at $M = M_{*}\sim10^{15}$g is mass below which PBHs would have Hawking evaporated by the current day (assuming no accretion or merging).  The line at $M \sim 3 M_\odot$ is the mass above which PBHs would be confused with astrophysical BHs.\label{fig_nu1}}
\end{figure}

\begin{figure}
\includegraphics[width=5.75in]{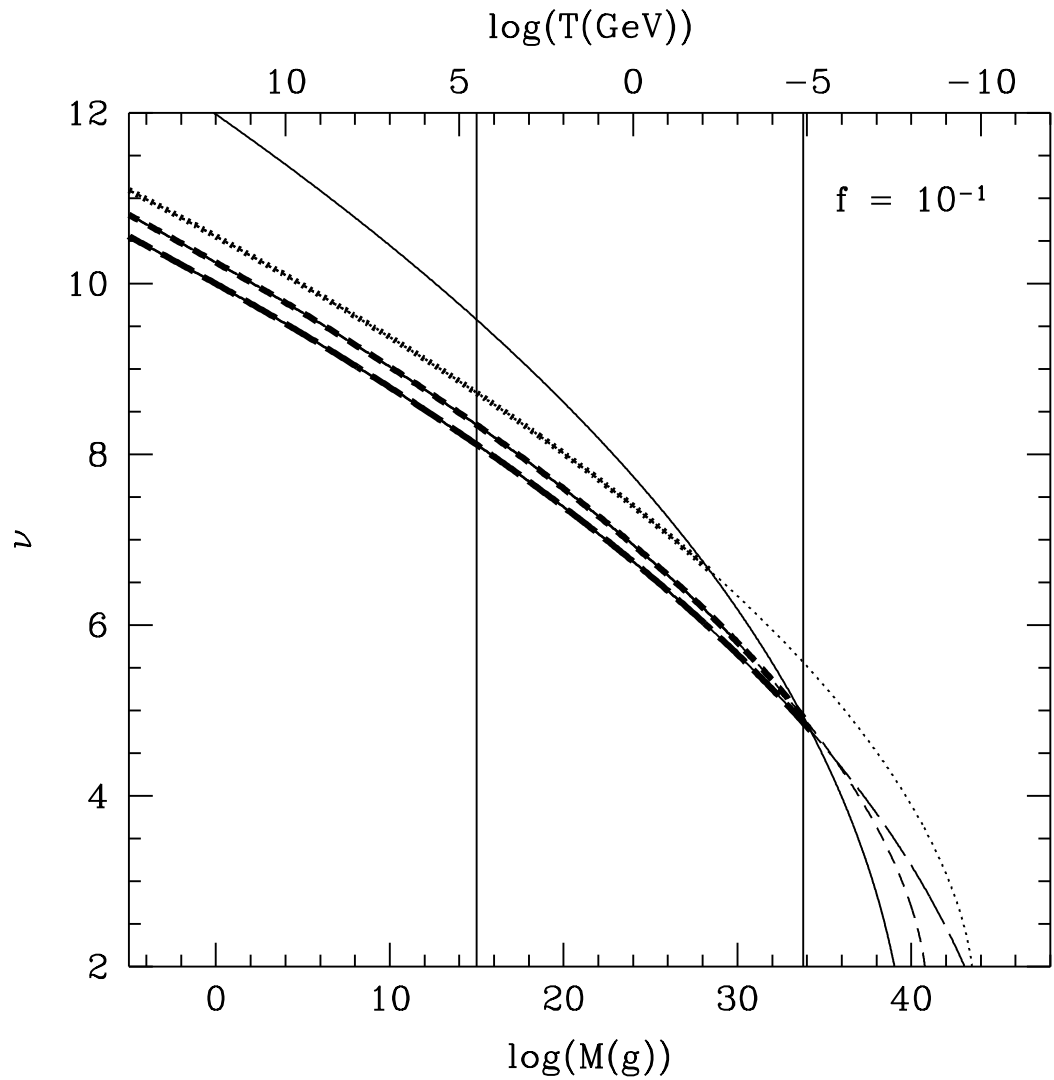}
\caption[Allowed region in $\nu-M_{PBH}$ space: $f=10^{-1}$]{The same as Figure~\ref{fig_nu1}, except with $f=0.1$.\label{fig_nu2}}
\end{figure}

\begin{figure}
\includegraphics[width=5.75in]{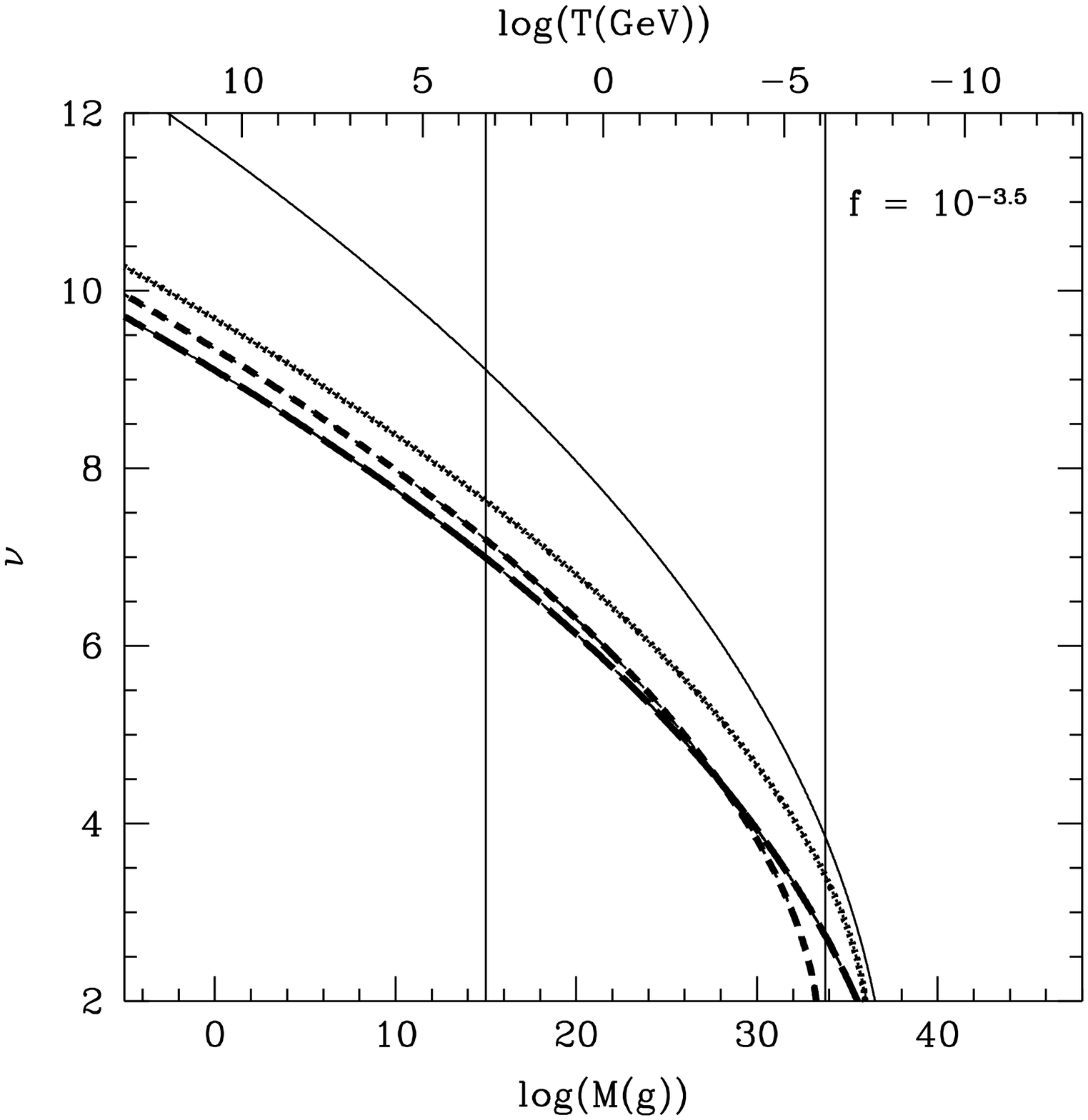}
\caption[Allowed region in $\nu-M_{PBH}$ space: $f=10^{-3.5}$]{The same as Figure~\ref{fig_nu1}, except with $f=10^{-3.5}$.\label{fig_nu3}}
\end{figure}

Having determined the initial PBH density, their abundance at subsequent times is simple to calculate.  PBHs 
are non-relativistic matter, so $\rho_{PBH} \propto a^{-3}$.  Because radiation redshifts as $\rho_{rad} \propto a^{-4}$,
the PBH to radiation ratio grows until the epoch of matter-radiation equality:
\begin{equation}
\frac{\rho_{PBH}}{\rho_{rad}}(t_{eq}) = \frac{B(t)}{1-B(t)} \left(\frac{t_{eq}}{t}\right)^{1/2}
\end{equation}
After the epoch of equality, $\Omega_{PBH}$ remains constant during matter domination up until the era of 
vacuum energy domination.  The condition that PBHs do not overclose the universe\footnote{This is also the condition that PBHs do not induce an early matter-dominated phase.} is $\Omega_{PBH}(t_{eq}) < 1/2$,
or
\begin{equation}
B(t) < \frac{1}{2} \left(\frac{t}{t_{eq}}\right)^{1/2}.
\end{equation}
Throughout, we will assume a monochromatic mass function such that $\rho_{PBH} = M n_{PBH}$.  We can then write
\begin{equation}
B(t) = f N_* e^{-\nu^2/2}.
\end{equation}
Figures~\ref{fig_nu1}~-~\ref{fig_nu3} show the lower limit on $\nu$ derived from the latter equations.
This exponential dependence of the PBH abundance upon $\nu$ means we must now then turn to a discussion 
of the form of the underlying power spectrum $P(k)$.  

A given mode crosses within the horizon at a time $t$ given by $k_H = a(t)H(t)$.  The radiation fluctuation 
on the horizon scale (crossing during radiation domination) is computed using Equation (\ref{sigmasquared})
\begin{equation}
\sigma_{rad}^2(r_H,t) = \frac{V}{2\pi^2} \int dk k^2 P_{rad}(k) T^2_{ad}(k,t) W^2_k(k r_H)
\end{equation}
For adiabatic perturbations (which we are assuming for radiation field), $T_{ad}(k,t) \propto k_H^{-2}$ (up to horizon 
crossing) and for a power law spectrum $P_{rad}(k) \propto k^n$, 
\begin{equation}\label{sigmascaling}
\sigma_{rad}^2(r_H) \propto k_H^{(n-1)}
\end{equation}
where $k_H = 1/r_H$.  The spectrum for $n=1$ is known as the Harrison-Zel'dovich \cite{harrison, zeldovich2}
spectrum (also called a scale-invariant spectrum) and corresponds to fluctuations of different physical sizes
having identical power when they enter the horizon.  Spectra with $n > 1$ are known as ``blue'' spectra, and 
correspond to models having more power at smaller scales (larger $k$).

During radiation domination, the horizon mass $M_H \propto t \propto a^2 \propto k_H^{-2}$, or $k_H \propto M_H^{-1/2}$, so that $\sigma_{rad}^2(r_H) \propto M_H^{(1-n)/2}$.  
During matter domination the scaling is different, $M_H \sim \rho R_H^3 \propto a^{-3} H^{-3} = k_H^{-3}$, or 
$k_H \propto M_H^{-1/3}$, so that $\sigma_{rad}^2 \propto M_H^{(1-n)/3}$.  For a pure power law spectrum then, we can
relate the power at any earlier time to the power today:
\begin{equation}
\sigma^2(r_H) = \sigma^2(H_{0}^{-1}) \left(\frac{M_{eq}}{M_0}\right)^{(1-n)/3} \left(\frac{M_H}{M_{eq}}\right)^{(1-n)/2}
\end{equation}
where $0$ subscripts refer to current values and $eq$ refers to the epoch of matter-radiation equality.  
From this, a value of $n>1$ can produce sufficient power at small 
scales to produce significant black holes.  Our understanding of the
physics at these scales in the early universe is only theoretical, and thus there may be
significant deviations from pure power-law behavior then.  

Due to quantum effects\cite{hawking2}, a BH of mass $M$ will emit particles as a blackbody 
with temperature $T_{h}$ given by
\begin{equation}
T_{h}(M) = \frac{1}{8\pi G M} = \frac{M_P^2}{8 \pi M} \approx 10^{22} \left(\frac{M}{1 \textrm{g}}\right)^{-1} \textrm{eV}.
\end{equation}
As the temperature is inversely proportional to the mass, this is unobservable for a one 
solar mass (and higher) BH ($T_h(M_\odot) \approx 62$ nK), but cannot be neglected in the mass range
of PBHs.  This emission also corresponds to a mass loss for the PBH, 
\begin{equation}
\dot M = - L_h = - \sigma_{SB}^* T_h^4 (4 \pi R_s^2) = -\frac{\alpha(M)}{M^2},
\end{equation}
where $\sigma_{SB}^*$ is the effective Stefan-Boltzmann constant and is related to the effective number
of relativistic degrees of freedom in the emitted particles.  PBHs therefore have a finite
lifetime, after which they would have emitted their entire rest mass, given by
\begin{equation}\label{lifetime}
\tau = \frac{M_0^3}{3\alpha(M_0)} \approx (10^{-26} \textrm{s}) \left(\frac{M}{1 \textrm{g}}\right)^3.
\end{equation}
The variation of the parameter $\alpha$ with mass is not great, changing by a factor of 10 over at least 7 decades
 of mass \cite{halzen}.  As the lifetime scales with $M^3$, there is a threshold mass above which holes will not have evaporated by 
the present day ($t_0$).  This threshold mass $M_*$ is given by
\begin{equation}
M_* \approx (4\times10^{14} \textrm{g}) \left[\left(\frac{\alpha(M_*)}{6.94\times10^{25}\textrm{g}^3/\textrm{s}}\right)\left(\frac{t_0}{4.4\times10^7 \textrm{s}}\right)\right]^{1/3}.
\end{equation}
Given the uncertainties in $\alpha$ and $t_0$, a threshold mass of $M_* \sim 10^{15}$ g is typically quoted in the literature.

A large enough abundance of PBHs with $M \approx M_*$ will produce a number of observable effects
through their evaporation in the current day.  They would contribute to cosmic rays \cite{macgibbon3}, the
$\gamma$-ray background \cite{page3,halzen}, 511 keV emission due to positron annihilation in the galactic center 
\cite{okele} or be the cause of short duration gamma ray bursts \cite{green3, cline}.
Observations (or the lack thereof) of PBHs evaporating today depend critically upon not only the 
number density of PBHs present today $n_{PBH}(t_0)$, but also upon how clustered they are within
within the galaxy.  Assuming an isothermal halo model, the effective number density is $\zeta n_{PBH}(t_0)$
where $\zeta$ is the local density enhancement factor \cite{macgibbon3, halzen, page3} and ranges from
$10^5 - 10^7$.

PBHs with $M < M_*$ would have evaporated by the present day.  The main mechanism for ``observing'' PBHs in cosmology is through their
Hawking radiation.  In the absence of a direct detection, the main utility of PBHs is to 
set limits of PBH abundance at various times given a non-detection.  Though, PBHs have also been
invoked to explain baryogenesis \cite{barrow2}, reionization \cite{gibilisco} 
and as a solution to the magnetic monopole problem \cite{stojkovic, stojkovic2}.

Evaporating PBHs have their most dramatic effect during the era
of BBN, where Hawking radiation can alter light element abundances
\cite{lindley1}.  Therefore, the success of BBN implies an upper limit to the
number of PBHs evaporating at that time.  

Combining Equations
(\ref{horizonmass}), (\ref{pbh_horizon_mass_relation}) and (\ref{lifetime}) gives the
relation
\begin{equation}\label{taut}
\tau(t) = \frac{f^3 M_P^3}{3 \alpha} \left(\frac{t}{t_P}\right)^3,
\end{equation}
the lifetime $\tau$ of a PBH created at a time $t$.  What this allows one to do is use 
information from a ``late epoch'' (time $\tau$) to examine conditions at an ``early epoch'' 
(time $t \ll \tau$).  In the above example, $\tau \sim t_{BBN}$, and the limits on initial
PBH abundance from BBN imply $\beta < 10^{-16}$ for $M_{PBH}$ between $10^{9}$ g 
and $10^{15}$ g (see, {\it i.e.}, \cite{GreLid}).

This relation depends critically upon the PBH mass monotonically decreasing due to evaporation, and not
gaining mass in any way (accretion or merging).  Should this not be the case, the lifetime $\tau$
is no longer given by the initial PBH mass, and the link between late epoch and early epoch
is broken.  Instead, the energy in PBHs that would have evaporated away can now linger for
longer periods of time.  It was shown in \cite{carr8} that PBHs will not appreciably increase their mass through 
radiation accretion.  PBH merging then would be the dominant mechanism for (significant) mass growth in
the radiation dominated epoch.  Since $\tau \propto M^3$, the merging of two equal mass BHs will
result in a BH with a lifetime 8 times as long.  If this merging can continue, then there is a 
greater chance of PBHs produced in the early universe still existing today.

Depending on the epoch of PBH formation, there is reason to believe there would be merging occuring
before, say, the epoch of nucleosynthesis, which could skew limits obtained from using Equation~(\ref{taut}).
Assuming an unclustered population, PBH binaries can form in the radiation era and be a source of gravitational
waves today \cite{nakamura}.  Any PBH clustering will only enhance the formation of close PBH binaries (and possibly
of larger bound structures), and orbital decay will cause merging before evaporation can occur.

\section{Bias Model}\label{bias}

Measuring the two point function (or its Fourier transform, the power spectrum) of astrophysical
objects is a powerful tool in studying their clustering properties.  The physical interpretation of $\xi(r)$
is as follows.  The differential probability of finding two objects (galaxies, clusters, PBHs, etc.) in volume
$dV_1$ and $dV_2$, a distance $r$ apart is given by
\begin{equation}
dP = \rho^2 (1 + \xi(r)) dV_1 dV_2.
\end{equation}
The two point function then measures the excess probability (over random) of finding pairs with a separation $r$ (here
and throughout we use comoving distances).
A large (positive) value of $\xi$ implies a large amount of clustering (objects are preferentially close to each other), a 
negative value of $\xi$ implies anti-clustering (objects are preferentially far away).

It is important to note that the galaxy-galaxy correlation function $\xi_{gg}$ is not identical
to the underlying mass correlation function $\xi_{m}$; in other words, galaxies are not a perfect
tracer of mass.  Further, different types of objects which may act as tracers (quasar, galaxy clusters)
have different clustering properties.  Measurements of $\xi$ for clusters of galaxies showed that they were
more clustered than galaxies themselves by a factor of 10.  
\citet{kaiser} showed that this may be explained using what is now known as the peak-background split
model of bias: as clusters of galaxies form from higher peaks in the density field than galaxies, it
is natural that they be more clustered.  In the limit of large separation and large peaks, the bias is
given by
\begin{equation}\label{kaiserbias}
\xi_{peak}(r) = \frac{\nu^2}{\sigma^2} \xi(r).
\end{equation}

This can be roughly understood as follows.  Split the density field into a long wavelength and a short
wavelength component.  Next, consider a peak in just the long wavelength component (``background''); the physical density
field will consist of this component modulated by the short wavelength portion.  If the threshold for
gravitational collapse is close to the value of the background peak value, the physical field will 
cross this threshold a number of times in the vicinity of the peak.  The regions above threshold, therefore,
are preferentially found near the background peak.


The assumptions used are:

{\bf PBH creation is rare:} PBH formation occurs during radiation domination ($w=1/3$); and the
radiation perturbations are gaussian.  At creation, there will be at most one PBH per horizon volume,
and PBH formation at around the horizon mass.

{\bf Peaks Theory bias:} Since PBH formation is a threshold process, we can use
peaks theory \cite{BBKS} to determine the number density and correlation statistics.  While we only consider
the two-point function and power spectrum here, all higher order correlation functions can be derived
in a similar manner.

We now derive the bias for a population of PBHs formed at a single mass scale, compared to the underlying
radiation field.  For the overdensities of PBHs and radiation $\delta_{PBH}$ and $\delta_{r}$, we define their two 
point correlation functions
\begin{equation}
\xi_{PBH}(r) = \langle \delta_{PBH}(x) \delta_{PBH}(x+r) \rangle
\end{equation}
\begin{equation}
\xi_{rad}(r) = \langle \delta_{rad}(x) \delta_{rad}(x+r) \rangle
\end{equation}
and the bias parameter
\begin{equation}
\xi_{PBH}(r) = b(r)^2 \xi_{rad}(r)
\end{equation}
Where, in general, $b(r)$ is not a constant.  The averaging done in the definition of the correlation functions includes
a window function on the scale of the horizon for smoothing.  Thus, the size of the fluctuations\footnote{While the terms {\it perturbation} and {\it fluctuation} are sometimes used interchangably in the literature to
refer to an inhomogeneity, we will make a distinction in the usage for radiation and PBHs.  The word 
{\it perturbation} typically implies smallness in the context of (cosmological) perturbation theory, and we use it to describe
the (initial) radiation field, as they will be no larger than order unity.  As we will show, this will not be the case for PBHs, and 
therefore we use the word {\it fluctuation} for their case.} 
in either the radiation and PBHs is characterized by
\begin{equation}
\sigma_{X,0}^2 = \xi_{X}(0) 
\end{equation}
From the definition of the radiation and PBH correlation functions, this is given by
\begin{equation}\label{bias2}
\sigma_{PBH,0} = b(0) \sigma_{rad,0}
\end{equation}
The power spectrum $P(k)$ is defined as
\begin{equation}
\label{def_ps}
P(k) = \left(\frac{4\pi}{V}\right) \int dr r^2 \xi(r) \left(\frac{\sin(kr)}{kr}\right).
\end{equation}

As PBHs form in regions above a certain threshold density, it is straight-forward to compute the number density
and bias assuming assuming PBHs form at a single mass only.  The bias is given by an integral over a bivariate
gaussian distribution; using the notation of \citet{jensen}, the full expression is given by
\begin{widetext}
\begin{eqnarray}
\label{jensen1}
1 + \xi_{PBH}(r) & = & \left[\frac{1}{2} \textrm{erfc}\left(\frac{\nu}{\sqrt{2}}\right)   \right]^{-2} \int_{\nu}^{\infty} dy_1 \int_{\nu}^{\infty} dy_2 (2\pi)^{-1} \left(1-w(r)^2\right)^{-1/2} \\ \nonumber
                 &   & \times \exp \left[- \frac{y_1^2 + y_2^2 - 2 y_1 y_2 w(r)}{2(1-w(r)^2)}\right]
\end{eqnarray}
\end{widetext}
where $w(r) = \xi_{rad}(r)/\sigma_{rad,0}^2$ is the normalized radiation correlation function and $\nu = \delta_c / \sigma_{rad,0}$.  
It is possible to write this as a power 
series  (the so-called tetrachoric series) in $w(r)$ \cite{jensen},
\begin{equation}
\label{jensen2}
\xi_{PBH}(r) = \sum_{m=1}^{\infty} \frac{A_m^2}{m!} w(r)^m
\end{equation}
where the coefficients are given by
\begin{equation}
A_m = \frac{2 H_{m-1}\left(\frac{\nu}{\sqrt{2}}\right)2^{-m/2}}{\sqrt{\pi} e^{\nu^2/2} \textrm{erfc}\left(\frac{\nu}{\sqrt{2}}\right)}
\end{equation}
where $H_n$ are the Hermite polynomials.

The result of \citet{kaiser} is obtained by assuming $w(r) \ll 1$ and $\nu \gg 1$, so that only the first term in the
series need be used to obtain $\xi_{PBH}(r) \approx \nu^2 w(r)$.  Relaxing the condition on $w(r)$ (but not on $\nu$),
the coefficients $A_m \rightarrow \nu^m$, obtaining the result of \citet{PolWis},
\begin{equation}
\label{politzer}
1 + \xi_{PBH}(r) = 1 + \sum_{m=1}^{\infty} \frac{(\nu^m)^2}{m!} w(r)^m = \exp(\nu^2 w(r))
\end{equation}

As $r \rightarrow 0$, $w \rightarrow 1$ by definition, so that in the case of arbitrary $\nu$,
\begin{equation}
\xi_{PBH}(0) = \left[\sqrt{\pi} e^{\nu^2/2} \textrm{erfc}\left(\frac{\nu}{\sqrt{2}}\right)\right]^{-2} \sum_{m=1}^{\infty} \frac{\left(2H_{m-1}\left(\frac{\nu}{\sqrt{2}}\right)\right)^2}{2^m m!}
\end{equation}

Recall that $\sigma^2_{PBH,0} = \xi_{PBH}(0)$.  For large $\nu$, it follows that $\sigma^2_{PBH,0} = e^{\nu^2}$.
In other words, PBHs start with a large fluctuation amplitude (compared to radiation) and their evolution 
begins in the nonlinear regime.  However, the number density goes as $e^{-\nu^2/2}$, so the fewer PBHs
are formed, the more clustered they will be.

Note this bias is independent of the PBH Mass - Horizon Mass relation (Equation~(\ref{pbh_horizon_mass_relation})).  
Specifically, we have computed
the correlation function of horizon sized regions that contain at most one PBH.  As such $P_{PBH}(k)$ will
have an initial upper cutoff at $k_H$.

While we can compute exactly $\xi_{PBH}$ from the peak-background split model, it is customary
in LSS studies to measure the power spectrum $P_{PBH}$ instead.  Inserting Equation $\ref{politzer}$
into Equation \ref{def_ps} we obtain the integral expression
\begin{eqnarray}
\label{P_PBH}
P_{PBH}(k) & = & \frac{4\pi}{V} \int dr r^2 \xi_{PBH}(r) \left(\frac{\sin (kr)}{kr}\right)\\ \nonumber
           & = & \frac{4\pi}{V} \int_{r_H}^{\infty} dr r^2 \left[\exp\left(\frac{\nu^2}{\sigma_{rad,0}^2}\xi_{rad}(r)\right) - 1\right]\left(\frac{\sin (kr)}{kr}\right)
\end{eqnarray}

The lower cutoff at $r_H = R_H/a = k_H^{-1}$, the comoving horizon length at PBH formation, is due to the 
finite size of the PBHs.  This will translate into an upper cutoff in $P_{PBH}(k)$ at $k_H$.
Generically, the above integral can be done numerically, but we can say more
about the nature of the PBH fluctuations without it.

By expanding the exponential, we can rewrite Equation~(\ref{P_PBH}) as\begin{widetext}
\begin{equation}
P_{PBH}(k) = \frac{\nu^2}{\sigma_{rad,0}^2} P_{rad} (k) + \sum_{m=2}^{\infty} \frac{4\pi}{V} \int_{r_H}^{\infty} dr r^2 \left(\frac{\sin (kr)}{kr}\right) \frac{1}{m!}\left[\frac{\nu^2}{\sigma_{rad,0}^2}\xi_{rad}(r)\right]^m
\end{equation}
\end{widetext}
The higher order terms in the above expansion show the non-linear dependence of $P_{PBH}$ upon $P_{rad}$.

Due to the discrete nature of the PBHs, the normalization condition for $P_{PBH}$ is that as $k \rightarrow 0$,
$P_{PBH}$ approaches a spectrum for pure Poisson noise; i.e., a constant value.  This is manifest in our above
expression.  The first term, where the PBH power spectrum is simply $b^2 P_{rad}$, with the bias $b$ given by the 
Kaiser value of $\nu/\sigma$.  We can Taylor expand the sine term in the integrals
such that $\sin(kr)/(kr) \rightarrow 1$, and those integrals evaluate to constants:
\begin{equation}
P_{Poisson} = \frac{1}{V} \sum_{m=2}^{\infty} \frac{4\pi}{m!} \frac{\nu^{2m}}{\sigma_{rad,0}^{2m}} \int_{r_H}^{\infty} dr r^2 \xi_r(r)^m
\end{equation}

The total PBH power spectrum then can be written as:
\begin{equation}\label{splitup}
P_{PBH}(k) = P_{Poisson} + \frac{\nu^2}{\sigma_{rad,0}^2} P_{rad} (k) + P_{SS} (k)
\end{equation}
where
\begin{equation}
P_{SS}(k) = \frac{1}{V} \sum_{l=1}^{\infty} \sum_{m=2}^{\infty} \int_{r_H}^{\infty} dr r^2 \left(\frac{(-1)^l(kr)^{2l}}{(2l+1)!}\right) \frac{4\pi}{m!}\left[\frac{\nu^2}{\sigma_{rad,0}^2}\xi_{rad}(r)\right]^m
\end{equation}
represents the small-scale power when $kr$ is not small.

\begin{figure}
\includegraphics[width=5.75in]{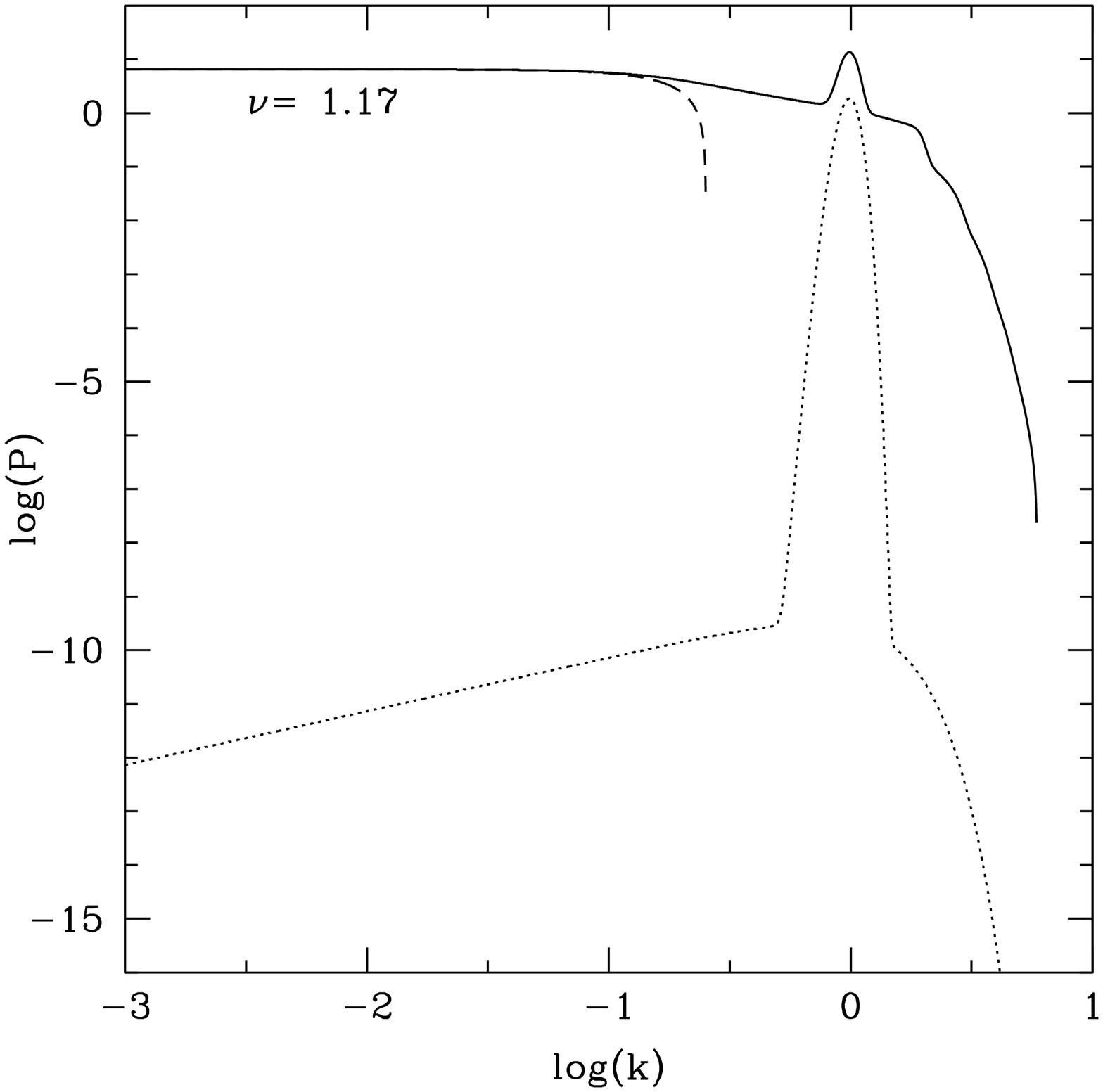}
\caption[PBH Power Spectrum: $\nu = 1.17$]{The PBH Power Spectrum for $\nu=1.17$.  Dotted line is the radiation power spectrum, consisting of a $n=1$ spectrum with COBE normalization, along with a gaussian spike in power at $k=1$.  Solid line is the PBH power spectrum, dashed line is the quadratic estimate of the PBH power spectrum. \label{power4}}
\end{figure}

\begin{figure}
\includegraphics[width=5.75in]{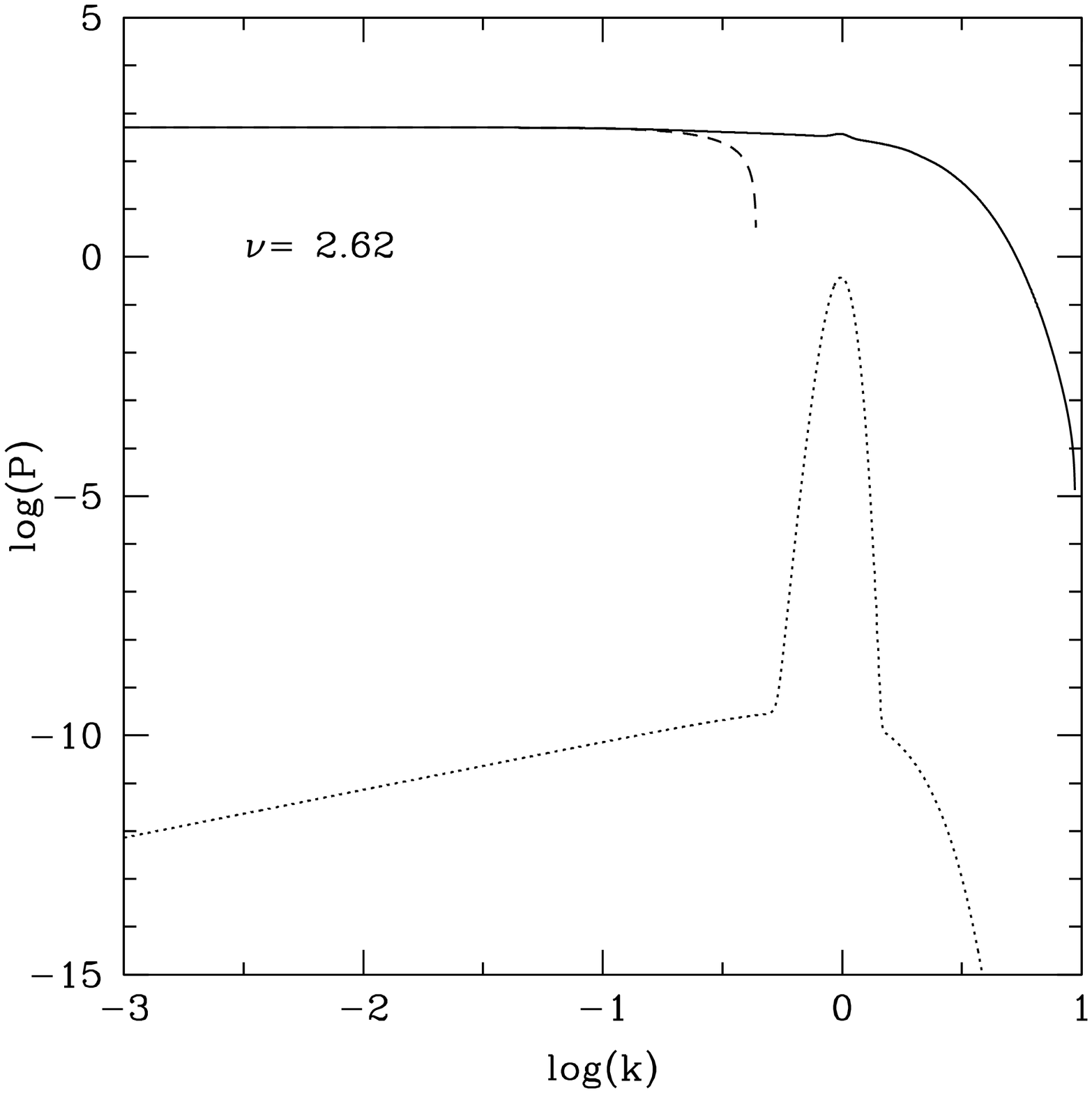}
\caption[PBH Power Spectrum: $\nu = 2.62$]{The PBH Power Spectrum for $\nu=2.62$.  Dotted line is the radiation power spectrum, consisting of a $n=1$ spectrum with COBE normalization, along with a gaussian spike in power at $k=1$.  Solid line is the PBH power spectrum, dashed line is the quadratic estimate of the PBH power spectrum. \label{power3}}
\end{figure}

\begin{figure}
\includegraphics[width=5.75in]{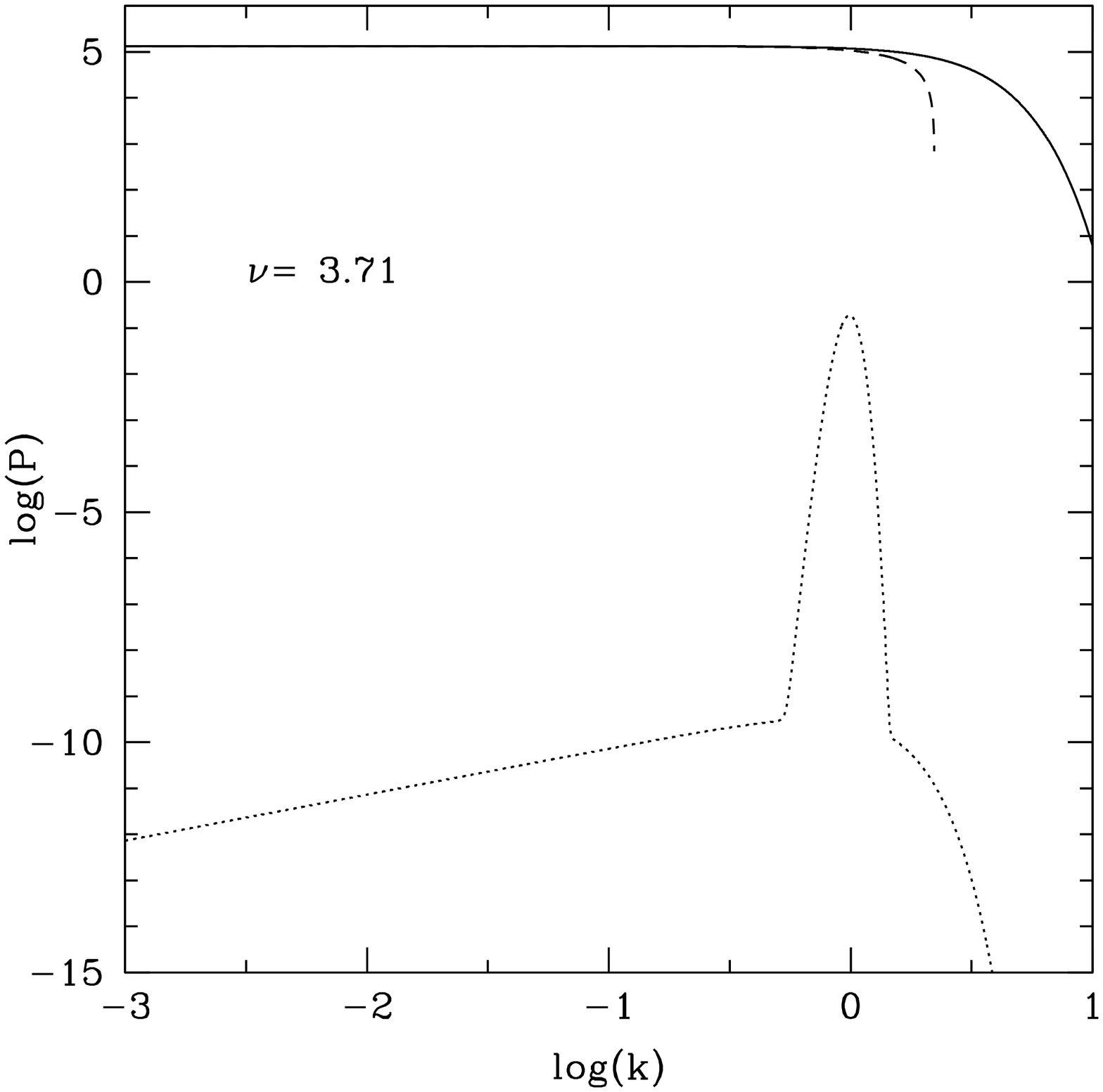}
\caption[PBH Power Spectrum: $\nu = 3.71$]{The PBH Power Spectrum for $\nu=3.71$.  Dotted line is the radiation power spectrum, consisting of a $n=1$ spectrum with COBE normalization, along with a gaussian spike in power at $k=1$.  Solid line is the PBH power spectrum, dashed line is the quadratic estimate of the PBH power spectrum. \label{power1}}
\end{figure}

\begin{figure}
\includegraphics[width=5.75in]{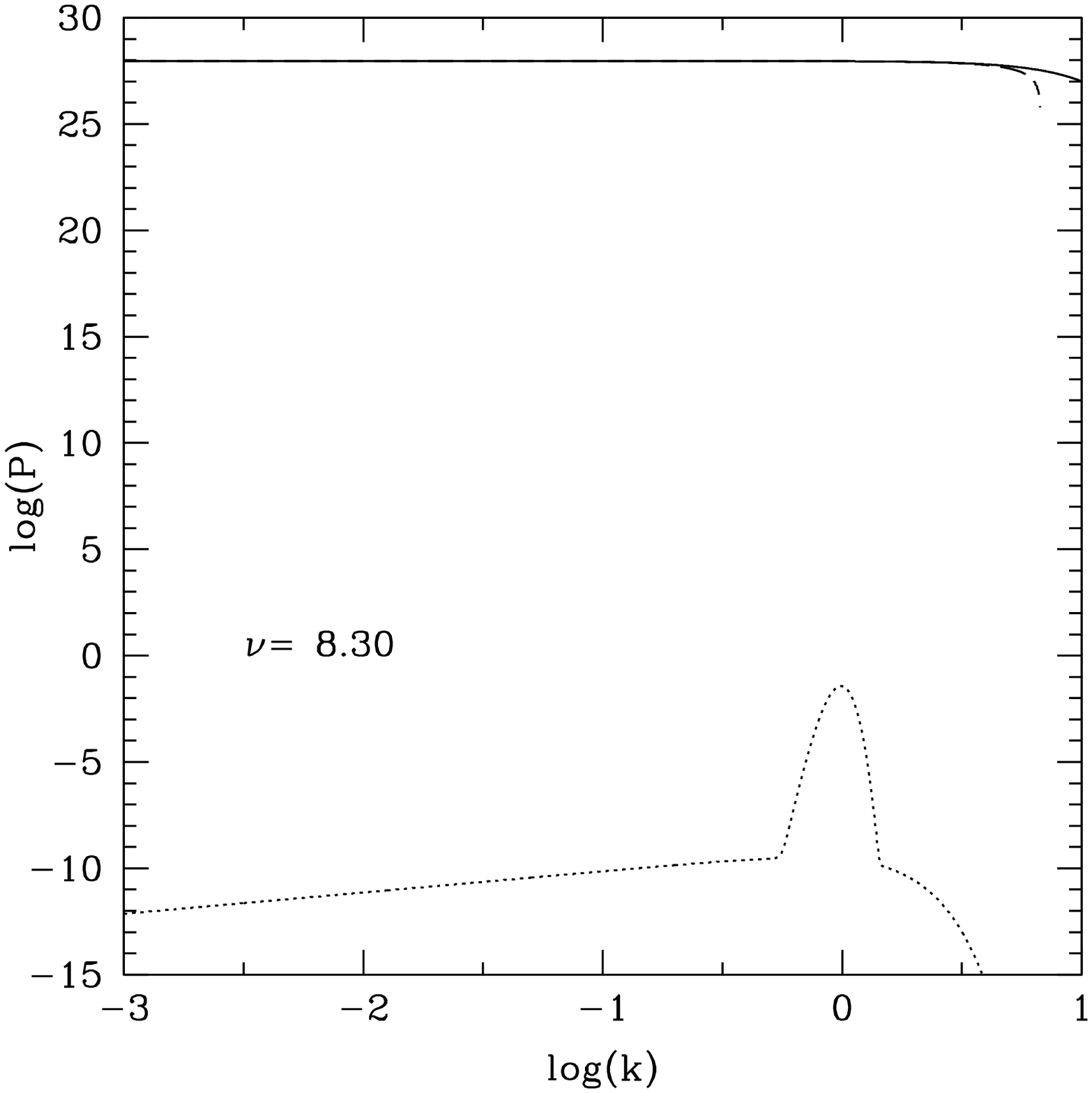}
\caption[PBH Power Spectrum: $\nu = 8.30$]{The PBH Power Spectrum for $\nu=8.30$.  Dotted line is the radiation power spectrum, consisting of a $n=1$ spectrum with COBE normalization, along with a gaussian spike in power at $k=1$.  Solid line is the PBH power spectrum, dashed line is the quadratic estimate of the PBH power spectrum. \label{power2}}
\end{figure}

\begin{figure}
\includegraphics[width=5.75in]{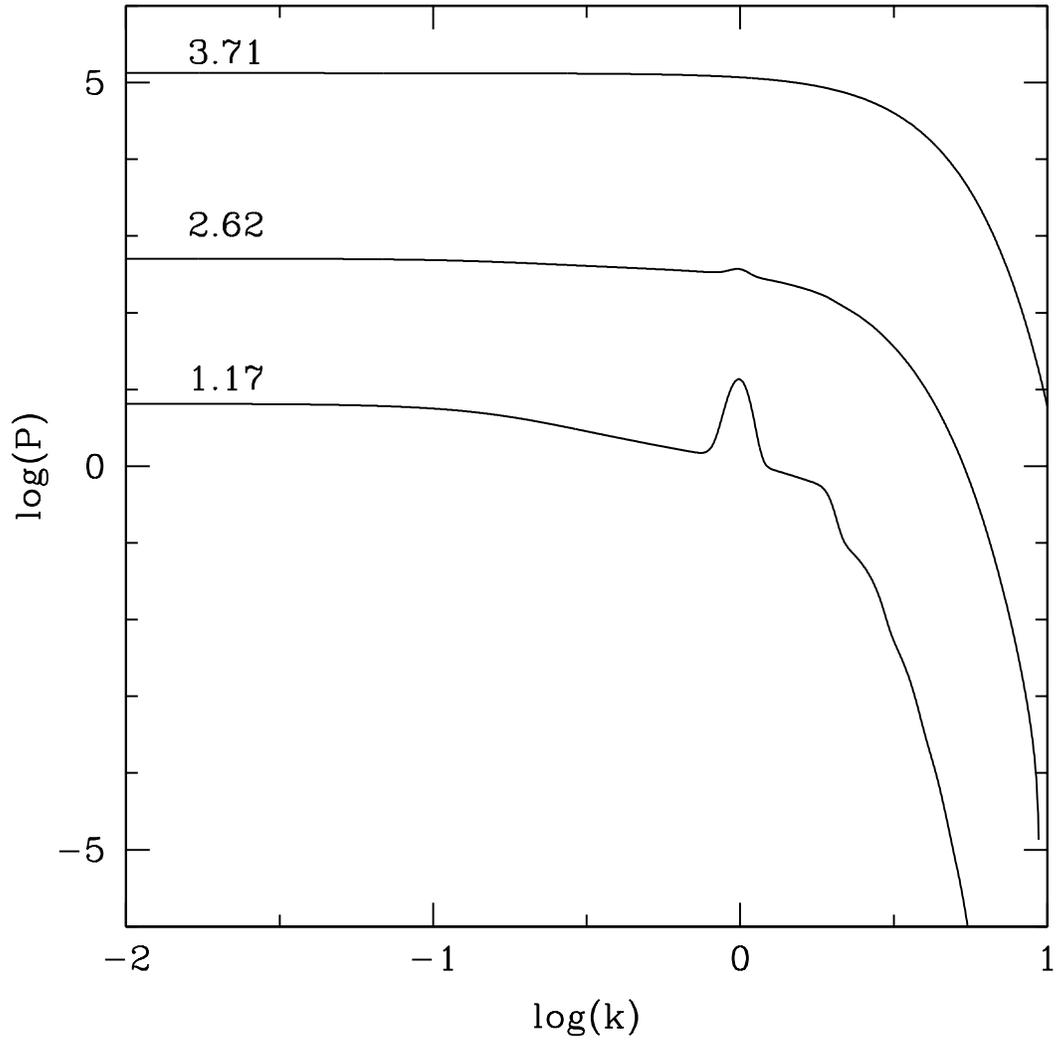}
\caption[PBH Power Spectrum:  Closeup]{The PBH Power Spectrum for $\nu=1.17, 2.62, 3.71$.   \label{totalpower}}
\end{figure}

To see the behavior of $P_{PBH}$ at small $k$, we numerically integrate Equation~(\ref{P_PBH}).  The 
underlying radiation power spectrum $P_{rad}$  is a $n=1$ spectrum normalized to the four-year Cosmic Background Explorer (COBE) value along with a 
gaussian spike at the horizon scale (the latter being normalized to unity).  For a fixed $\delta_c = 2/3$, varying
the spike amplitude will vary the value of $\nu$.  Figures \ref{power4} to \ref{totalpower} show $P_{PBH}(k)$ for
four values of $\nu$.  We see that as $\nu$ increases, the constant (Poisson) power quickly damps out the 
linear (Kaiser) term.  The $l=2$ terms of $P_{SS}(k)$ survive for intermediate values of $k$ as a small negative quadratic
contribution ($\propto -k^2$).

Note that the power spectrum given in Equation~(\ref{splitup}) is the initial spectrum immediately after PBH creation.  
Due to the different $k$-dependence of each of the terms, the power at later times ($k \ll k_H$) will not
be dominated by the Poisson term.  We return to this in Section~\ref{PBHfluctuationevolution} where we compute the power
at horizon crossing at later times.

From Equation~(\ref{P_PBH}), 
we expect $P_{Poisson} \sim e^{\nu^2}$; a better fit for $\nu \gtrsim 4$ yields
\begin{equation}
P_{Poisson} \approx \frac{10}{7} \exp \left(\frac{3}{4} \nu^{2.1}\right).
\end{equation}
The power spectrum for a group of $N$ objects randomly distributed (with a uniform distribution) 
is $1/N = (nV)^{-1} = \beta^{-1}$.  Note that our above expression for $P_{Poisson} \neq \beta^{-1}$, indicating
the PBHs are distributed as clusters of objects with mean occupation number
\begin{eqnarray}\label{N_c}
N_c & = & P_{Poisson} \beta \\ \nonumber
       & = & \frac{10}{7} N_*(\nu) \exp \left(\frac{3}{4} \nu^{2.1} - \frac{1}{2}\nu^2\right) \\ \nonumber
       & \sim & N_*(\nu) e^{\nu^2/4}.
\end{eqnarray}

\subsection{Adiabatic vs. Isocurvature}

We now take an aside to further consider the nature of the PBH fluctuations.  
That PBHs correspond to isocurvature perturbations has been noted in the literature \cite{carr5,carrsilk,BBKS,freese}, 
though it has not received a lot of attention in the recent PBH publications.  In models where PBHs
constitute the dark matter, it was assumed that their perturbations would be purely adiabatic, as with other
types of dark matter.  We point out that this is not the case; a large isocurvature component exists at shorter scales
in addition to the adiabatic component at longer scales.

To demonstrate this, assume that radiation is the only component in the universe; there is, therefore,
no distinction between adiabatic or isocurvature type perturbations.  The radiation perturbation 
corresponds to a perturbation in the spatial curvature\footnote{Whether the perturbation is Gaussian or
non-Gaussian is largely irrelevent at this point; perturbations of order unity must be non-Gaussian to some degree,
and we will show in the next section that the perturbations of PBHs are generically non-Gaussian.}.
Once PBHs are created from gravitational collapse, they will evolve as a matter ($w=0$) field in the
universe.  As such, we can examine the fluctuations in the PBH density.
At the time of PBH creation, {\it their} fluctuations can be classified as either adiabatic
or isocurvature.  By assumption, PBHs form from the collapse of a density perturbation once it enters the 
horizon.  In the radiation dominated era, the PBH mass is close to the horizon mass, so that at most
one PBH forms per horizon volume.  Each PBH is separated by at least a horizon distance.  The population
cannot have correlations on scales smaller than the horizon, so that the perturbations only exist for
super-horizon scales.  Any super-horizon perturbation can be written as a sum of adiabatic and 
isocurvature modes.

Note that, in our setup, only after PBH creation does the distinction between adiabatic and isocurvature
perturbations exist.  We intend to prove that the PBH fluctuations have an isocurvature component.  This 
can be generalized to the case where there are additional fields and the initial perturbation is wholly 
adiabatic.


Isocurvature perturbations correspond to perturbations in the local equation of state $w = p/\rho$, 
while adiabatic perturbations correspond to perturbations in the local energy density, and thus the
local curvature.  Consider a volume of space greater than the horizon volume at PBH creation.  The
formation of PBHs cannot change the energy density within this space: the gravitational collapse
corresponds to a ``shuffling" of energy density from one form (radiation) into another (matter).  The
decrement in the radiation energy density is exactly balanced by the creation of PBH energy density.
Therefore, the curvature is unchanged on super-horizon scales.  The total perturbation will only 
become adiabatic if this ``shuffling" takes place as to satisfy the adiabatic condition.
Further, by the second law of black hole thermodynamics, a black hole will always have a higher entropy than
the material that formed it.  PBH formation thus corresponds to an increase in entropy, and should this 
process occur non-uniformly, this will result in entropy perturbations, {\it i.e.} isocurvature perturbations.

The proof that the PBH fluctuations are isocurvature, then, derives from the fact that PBH formation is 
highly non-uniform.  Equivalently, that PBHs are created highly clustered, which was shown earlier in this 
section.  Using the notation of \cite{liddlelyth}, we write the entropy perturbation as
\begin{equation}\label{entropy}
S_{PBH} = \delta_{PBH} - \frac{3}{4}\delta'_{rad}
\end{equation}
where $\delta'_{rad}$ is the radiation perturbation {\it after} PBH formation, and $\delta'_{rad} \neq \delta_{rad}$.
Using the parameter $B$ from Equation~(\ref{omega}), we can trivially write
\begin{equation}
\rho_{rad} = \rho'_{rad} + \rho_{PBH} = (1-B)\rho_{rad} + B \rho_{rad}
\end{equation}
which allows us to relate the perturbations as
\begin{equation}
\delta_{rad} = (1-B) \delta'_{rad} + B \delta_{PBH}.
\end{equation}
We can use this latter equation to rewrite Equation~(\ref{entropy}) as
\begin{eqnarray}
S_{PBH} & = & \delta_{PBH}\left(1 + \frac{3}{4}\frac{B}{1-B}\right) - \frac{3}{4}\frac{1}{1-B} \delta_{rad}\\
              & \approx & \delta_{PBH} - \frac{3}{4} \delta_{rad}
\end{eqnarray}
which is now a function of the {\it initial} radiation perturbation and the (final) PBH fluctuation, and the approximation
holds as long as $B \ll 1$.  While $\delta_{rad} < 1$ by assumption, we know from Equation~(\ref{bias2}) that $\delta_{PBH}$
typically will not due to clustering.  We see that the entropy perturbation is simply a function of the bias parameter:
\begin{equation}
S_{PBH} \approx \left(b - \frac{3}{4}\right) \delta_{rad}.
\end{equation}
It is apparent that the isocurvature perturbation is almost inevitable for realistic (rare) PBH production.  The bias
parameter $b$ will be dependent on scale; in Fourier space $b$ is given roughly by $\sqrt{P_{PBH}/P_{rad}}$.
For a given $k$, the bias is dominated by the term domination the power spectrum as given in Equation~(\ref{splitup}).
At the smallest scales (close to PBH creation scales), the bias is largest and using Equation~(\ref{politzer}) gives
$b \sim \exp(\nu^2/2)/\sigma \gg 1$.  For larger scales, the linear (Kaiser) bias gives $b = \nu/\sigma$.  In either case,
the parameters ($\sigma, \nu$) would have to be finely tuned in order to produce a purely adiabatic PBH perturbation.

We note that this mechanism for generating an isocurvature perturbation is independent of the process that
created the initial (adiabatic) perturbation, though we assume throughout that it is done through an epoch of cosmological
inflation.  This
mechanism then is an exception to the generally held thought that an isocurvature perturbation cannot be produced
from single field inflation \cite{weinberg2004}.  The reason this occurs is that PBH creation ({\it i.e.} gravitational collapse) 
is an inherently non-linear and non-perturbative process that is not bound by this restriction from perturbation theory.  PBH
dark matter is not like particulate dark matter.  Further, for PBHs lighter than $M_*$ this isocurvature fluctuation is transferred
to the products of Hawking evaporation.  Thus, the absence of an observed isocurvature perturbation implies a limit on the number
of PBHs that have evaporated in the past.  We plan to further explore this topic in a future paper. 

\subsection{Gaussian vs. Non-Gaussian}

In our derivation of the PBH number density and clustering properties, we assumed the underlying radiation perturbation was
gaussian.  As PBHs form only at the peaks of the density field, and the initial size of the fluctuation is greater than unity,
the PBH fluctuations cannot be gaussian.  They appear instead to be lognormal (LN) in character.  Roughly, a LN distribution
is the exponentiation of a gaussian distribution.  The two-point correlation function of a LN field is given by \cite{coles}
\begin{equation}
1 + \xi_{LN}(r) = \exp\left(\Xi(r)\right),
\end{equation}
where $\Xi(r)$ is the correlation function of a gaussian field with variance $\Xi(0) = S^2$.  From Equation~(\ref{politzer}), 
this is exactly the correlation function for the PBH population assuming $\Xi(r) = \nu^2 w(r)$ and $\nu \gg 1$.

An isocurvature perturbation necessarily is defined between two components, here radiation and PBHs.  While we have
been focusing on the PBHs, there is of course a change in the radiation perturbations; the increase in PBH density 
is exactly cancelled by a decrease in radiation density.  From the perspective of the radiation field, not only is there 
an isocurvature component along with the (initial) adiabatic component, but there is now a non-gaussian fluctuation
along with the (initial) gaussian one.

\subsection{PBH Fluctuation evolution}\label{PBHfluctuationevolution}

The evolution of the PBH population after creation is a complex problem, outside the bounds of perturbation
theory due to the size of the initial PBH fluctuations, and better addressed as an N-body problem \cite{chisholm}.
However, we can make a rough estimate of the power at horizon crossing of other scales using the results from 
cosmological perturbation theory.  We can break the PBH power spectrum into its isocurvature and adiabatic
components:
\begin{eqnarray}
P_{PBH}(k) & = & \left(P_{PBH}(k) - \frac{9}{16}P_{rad}(k)\right) + \frac{9}{16}P_{rad}(k) \nonumber \\
		     & = & P_{iso}(k) + P_{ad}(k).
\end{eqnarray}
We can then write the variance at horizon crossing as
\begin{equation}
\sigma_{PBH}^2(r_H,t) = \frac{V}{2\pi^2} \int dk k^2 \left(P_{iso}(k) T^2_{iso}(k,t) + P_{ad}(k)T^2_{ad}(k,t)\right)W^2_k(k r_H).
\end{equation}

Rather than computing this explicitly, we will note that for power law spectra where $P_{iso}(k)\propto k^{n_{iso}}$ and
$P_{ad}(k)\propto k^n$, their contributions to the variance at horizon crossing can be written as
\begin{eqnarray}
\sigma_{ad}^2(r_H) & = & \sigma_{ad}^2(H_0^{-1})\left(\frac{M_{eq}}{M_0}\right)^{\frac{1-n}{3}} \left(\frac{M_H}{M_{eq}}\right)^{\frac{1-n}{2}}, \\ \nonumber
\sigma_{iso}^2(r_H) & = & \sigma_{iso}^2(H_0^{-1})\left(\frac{M_{eq}}{M_0}\right)^{\frac{(n_{iso}+3)}{3}} \left(\frac{M_H}{M_{eq}}\right)^{\frac{(n_{iso}+3)}{2}}, \\
\end{eqnarray}
while the total variance is their sum:
\begin{equation}
\sigma^2 = \sigma_{iso}^2 + \sigma_{ad}^2.
\end{equation}

The condition for scale-invariance is no scaling with mass; for adiabatic perturbations this is $n=1$, for isocurvature 
perturbations this is $n_{iso}=-3$.  While the adiabatic portion of $P_{PBH}(k)$ is scale-invariant by assumption, for scales
larger than the horizon size at their creation, the isocurvature component has a flat spectrum ($n_{iso} \approx 0$) and 
diminishes at longer scales.  Thus while the isocurvature portion dominates initially, there is a crossover scale where the 
spectrum becomes adiabatic.  Given the lack of measured isocurvature component at the time of the CMB (upper limit on 
isocurvature fraction is $f_{iso} < 0.33$, from \cite{peiris}), we can put a limit on the PBH population so that it does not violate
this bound.  Roughly, at the scale of matter-radiation equality ($M_{EQ} \sim 10^{48}$g), 
\begin{equation}
\sigma^2(r_{EQ}) = \sigma_{iso}^2 + \sigma_{ad}^2 =  \delta_H^2,
\end{equation}
and the bound is
\begin{equation}\label{fluclimit}
\sigma^2_{iso}(r_{EQ}) < f^2_{iso} \delta_H^2,
\end{equation}
where $\delta_H = 1.91\times10^{-5}$.
To compute $\sigma^2_{iso}(r_{EQ})$, we assume $P_{iso}(k) \approx P_{Poisson}$, which, as shown in 
Figures~\ref{power4}~-~\ref{power2}, is valid for $k \lesssim k_H/10$.  
The upper limit is plotted in Figures~\ref{fig_nu1}~-~\ref{fig_nu3} for three different values of $f$.  This constraint
becomes an upper limits on the (initial) PBH mass if it is to serve as dark matter.  For $f=1$, allowed regions for PBH dark 
matter all have $M_{PBH} < M_\odot$, so that there would be no confusion with astrophysical BHs.  As we decrease $f$, the 
upper limit increases: for $f=10^{-3.5}$ it is pushed above the confusion limit.

\section{Conclusions}\label{conc}

We have shown that for PBHs to serve as dark matter, clustering constrains them to lie in a 
particular mass range.  Further, PBHs will preferentially be found in clusters.

As shown in the previous section, PBH fluctuations enter the horizon with a very large
amplitude ($\sigma_{PBH} \sim e^{\nu^2/2}$).  It is therefore no longer value to treat
their evolution using linear perturbation theory, as one is able to do for other forms
of CDM.  Instead, we examine the sub-horizon evolution of the PBH population as an N-body
problem.  Being non-relativistic, PBHs will 
cluster hierarchically (just as CDM); creating smaller bound systems that get incorporated
into larger ones.  The internal dynamics of these systems are determined solely by gravitational
clustering, analogous to other gravitationally bound systems such as star clusters and galaxies.
For this, we are aided by the work done in the context of
studying more massive black holes in globular clusters \cite{sigurdsson} and galaxies \cite{begelman}.
In those cases, gravitational interactions tend to either produce bound pairs or ejections,
rather than BH coalescence \cite{saslaw}.  

What occurs in the case of PBHs depends upon
how many form in a ``PBH cluster'' and what their initial separations are.  The estimate of cluster population
in Equation~(\ref{N_c}) is likely an overestimate since our approximation for $\xi_{PBH}(r)$ breaks down for small $r$.
The initial separations should be on order the horizon size at formation, being the only length scale involved
in PBH formation.  This would seem to indicate rather compact clusters (initial separation on order the size of the PBHs
themselves), though more work ({\it e.g.}, higher order statistics, numerical simulations) is needed to verify this.

Frequent merging due to clustering could have a profound impact upon cosmology.  Since their
lifetime $\tau \propto M^3$, PBHs, due to merging, exist longer than they would have initially.  This 
could feasibly lead to a PBH population in the present universe that was formed in the earliest
moments of the early universe, opening up a new and unique observational window into that time.
At the very least, PBH merging in clusters dramatically changes the limits on initial PBH 
abundance, such as those used to put limits on models of inflation\citep{carr10,carr2,kim1,GreLid, green1,kim2,kribs,he,bugaev}.  
The issue of PBH clusters and merging is discussed more fully in a companion paper \citep{chisholm}.

Limits on the current number density of PBHs depend critically upon how clustered
PBHs are in our galaxy.  Naively, from our work in this paper, we might expect a local clustering 
enhancement $\zeta \sim e^{\nu^2/2}$, or $\zeta \sim 10^{22}$ for $\nu=10$.  This is many orders of
magnitude larger than the factors of $10^7$ computed in the literature.  This ignores the effect of
PBH merging though; sufficient merging might concentrate all galactic PBHs into the center SMBH.
This will have implications for models where PBHs are used to be the ``seed" BHs needed for the
growth of SMBHs in the centers of galaxies \cite{bean, duchting, custodio1}.

This PBH merging scenario we have discussed has
other predictions.  One prediction is more gravitational wave emission than originally assumed
for a uniform PBH population.  This is due to the increased probability of PBH binary formation
and emission from resonant bound states.

The other prediction is related to dark matter. 
Suppose now that PBHs are not the only component of the dark matter, and that there also exists
a ``standard'' CDM candidate with adiabatic perturbations (in accordance with CMB measurements),
in which case the CDM perturbation amplitude is related to the radiation perturbation amplitude
by $\delta_{CDM} = (3/4) \delta_{rad}$. 

Perturbations in the radiation density can only collapse (into PBHs) if they are of sufficient
amplitude on the scale of the horizon.  Perturbations smaller than this, in accordance with
linear perturbation theory, will simply oscillate, but not collapse.  This implies that there will be scales
slightly larger than those where PBH formation took place where $\delta_r$ is below the threshold
for PBH formation but still large compared to, say, the amplitude at the time of the CMB ($10^{-5}$).  
There is, accordingly, a similarly large perturbation in the CDM density assuming adiabaticity.  
While the linear growth of matter perturbations is delayed until after matter-radiation equality,
they still grow logarithmically in the radiation dominated era.  This leads to the possibility that they will become
non-linear before equality, and forming bound dark matter structures along with PBHs.  In which case,
one would have to include the interaction between these two populations of primordial bound objects.

\begin{acknowledgements}
The author would like to thank Rocky Kolb, John Carlstrom, Sean Carroll, Ilya Gruzberg, 
Robert Wald, Anne Green, Scott Dodelson, Jim Fry, David Wands and Niayesh Afshordi for helpful discussions and
feedback on this manuscript.  This work was supported in part by the Department of Energy.
\end{acknowledgements}

\clearpage

\clearpage

\clearpage

\clearpage

\clearpage

\clearpage

\clearpage

\clearpage


\clearpage

\begin{thebibliography}{99}
\bibitem[Zel'dovich \& Novikov(1967)]{zeldovich}
Ya. B. Zel'dovich, and I. D. Novikov, Sov. Astron. A. J. 10, 602 (1967)

\bibitem[Hawking(1971)]{hawking1}
S. W. Hawking, Mon. Not. R. Astron. Soc. 152, 75 (1971)

\bibitem[Polnarev \& Zembowicz(1991)]{polnarev}
A. Polnarev, and R. Zembowicz, Phys. Rev. D 43, 1106 (1991)

\bibitem[Rubin, Khlopov \& Sakharov(2000)]{rubin}
S. G. Rubin, M. Yu. Khlopov, and A. S. Sakharov, Grav. Cosm. S6, 51 (2000)

\bibitem[Hawking, Moss \& Stewart(1982)]{HMS}
S. W. Hawking, I. G. Moss, and J. M. Stewart, Phys. Rev. D 26, 2681 (1982)

\bibitem[Spergel et al.(2003)]{WMAP1}
D. N. Spergel et al., Ap. J. Supp. 148, 175 (2003)

\bibitem[Liddle \& Lyth(2000)]{liddlelyth}
A. R. Liddle, and D. H. Lyth, {\it Cosmological Inflation and Large-Scale Structure}, Cambridge University Press, Cambridge, UK (2000)

\bibitem[Chapline(1975)]{chapline}
C. F. Chapline, Nature 25, 251 (1975)

\bibitem[Ichiki, Orito \& Kajino(2003)]{ichiki}
K. Ichiki, M. Orito, and T. Kajino, Astropart. Phys. 20, 499 (2003)

\bibitem[Jedamzik(1997)]{jedamzik}
K. Jedamzik, Phys. Rev. D 55, 5871 (1997)

\bibitem[Green \& Jedamzik(2002)]{GreJed}
A. M. Green, and K. Jedamzik, Astron. Astrophys. 395, 31 (2002)

\bibitem[Blais, Kiefer \& Polarski(2002)]{blais}
D. Blais, C. Kiefer, and D. Polarski, Phys. Lett. B 535, 11 (2002)

\bibitem[Blais et al.(2003)]{blais2}
D. Blais, T. Bringmann, C. Kiefer, and D. Polarski, Phys. Rev. D 67, 024024 (2003)

\bibitem[Ivanov, Naselsky \& Novikov(1994)]{ivanov}
P. Ivanov, P. Naselsky, and I. Novikov, Phys. Rev. D 50, 7173 (1994)

\bibitem[Ivanov(1998)]{ivanov2}
P. Ivanov, Phys. Rev. D 57, 7145 (1998)

\bibitem[Yokoyama(1997)]{yokoyama3}
J. Yokoyama, Astron. Astrophys. 318, 673 (1997)

\bibitem[Bassett \& Tsujikawa(2001)]{bassett}
B. A. Bassett, and S. Tsujikawa, Phys. Rev. D 63, 123503 (2001)

\bibitem[Green \& Malik(2001)]{GreMal}
A. M. Green, and K. A. Malik, Phys. Rev. D 64, 021301 (2001)

\bibitem[Suyama et al.(2004)]{Suyama}
T. Suyama, T. Tanaka, B. Bassett, and H. Kudoh, hep-ph/0410247 (2004)

\bibitem[Barrau et al.(2004)]{barrau}
A. Barrau, D. Blais, G. Boudoul, and D. Polarski, Ann. Phys. 13, 115 (2004)

\bibitem[Barrow, Copeland \& Liddle(1992)]{barrow}
J. D. Barrow, E. J. Copeland, and A. R. Liddle, Phys. Rev. D 46, 645 (1992)

\bibitem[MacGibbon(1987)]{macgibbon}
J. H. MacGibbon, Nature 329, 308 (1987)

\bibitem[Carr(2005)]{carrreview}
B. J. Carr, astro-ph/0511743 (2005)

\bibitem[M\'{e}sz\'{a}ros(1975)]{meszaros}
P. M\'{e}sz\'{a}ros, Astron. Astrophys. 38, 5 (1975)

\bibitem[Carr(1975)]{carr5}
B. J. Carr, Ap. J. 201, 1 (1975)

\bibitem[Carr(1977b)]{carr11}
B. J. Carr, Astron. Astrophys. 56, 377 (1977)

\bibitem[M\'{e}sz\'{a}ros(1980)]{meszaros2}
P. M\'{e}sz\'{a}ros, Ap. J. 238, 781 (1980)

\bibitem[Kotok \& Naselsky(1998)]{kotok}
E. Kotok, and P. Naselsky, Phys. Rev. D 58, 103517 (1998)

\bibitem[Garc\'{\i}a-Bellido, Linde \& Wands(1996)]{garcia}
J. Garc\'{\i}a-Bellido, A. Linde, and D. Wands, Phys. Rev. D 54, 6040 (1996)

\bibitem[Garc\'{\i}a-Bellido \& Linde(1998)]{garcia2}
J. Garc\'{\i}a-Bellido, and A. Linde, Phys. Rev. D 57, 6075 (1998)

\bibitem[Lemoine(2000)]{lemoine}
M. Lemoine, Phys. Lett. B 481, 333 (2000)

\bibitem[Green(1999)]{green2}
A. M. Green, Phys. Rev. D 60, 063516 (1999)

\bibitem[Khlopov \& Barrau(2004)]{khlopov}
M. Yu Khlopov, and A. Barrau, astro-ph/0406621 (2004)

\bibitem[Afshordi, McDonald \& Spergel(2003)]{afshordi}
N. Afshordi, P. McDonald, and D. N. Spergel, Ap. J. 594, L71 (2003) 

\bibitem[Freese, Price \& Schramm(1983)]{freese}
K. Freese, R. Price, and D. N. Schramm, Ap. J. 274, 405 (1983)

\bibitem[Hiscock(1998)]{hiscock}
W. A. Hiscock, Ap. J. Lett. 509, 101 (1998)

\bibitem[Ioka et al.(1998)]{ioka1}
K. Ioka, T. Chiba, T. Tanaka, and T. Nakamura, Phys. Rev. D 58, 063003 (1998)

\bibitem[Ioka, Tanaka \& Nakamura(1999)]{ioka2}
K. Ioka, T. Tanaka, and T. Nakamura, Phys. Rev. D 60, 083512 (1999)

\bibitem[Nakamura et al.(1997)]{nakamura}
T. Nakamura, M. Sasaki, T. Tanaka, and K. S. Thorne, Ap. J. 487, L139 (1997)

\bibitem[Abbott et al.(2005)]{abbott}
B. Abbott et al., gr-qc/0505042 (2005)

\bibitem[Carr \& Hawking(1974)]{carr8}
B. J. Carr, and S. W. Hawking, Mon. Not. R. Astron. Soc. 168, 399 (1974)

\bibitem[Hawke \& Stewart(2002)]{hawke}
I. Hawke, and J. M. Stewart, Class. Quantum Grav. 19, 3687 (2002)

\bibitem[Musco, Miller \& Rezzolla(2005)]{musco}
I. Musco, J. C. Miller, and L. Rezzolla, Class. Quantum Grav. 19, 1405 (2005)

\bibitem[Neimeyer \& Jedamzik(1999)]{niemeyer}
J. C. Niemeyer, and K. Jedamzik, Phys. Rev. D 59, 124013 (1999)


\bibitem[Bardeen et al.(1986)]{BBKS}
J. M. Bardeen, J. R. Bond, N. Kaiser, and A. S. Szalay, Ap. J. 304, 15 (1986)

\bibitem[Green et al.(2004)]{GLMS}
A. M. Green, A. R. Liddle, K. A. Malik, and M. Sasaki, Phys. Rev. D 70, 041502 (2004)

\bibitem[Green(2005)]{green_private_communication}
A. M. Green, private communication (2005)

\bibitem[Harrison(1970)]{harrison}
E. R. Harrison, Phys. Rev. D 1, 2726 (1970)

\bibitem[Zel'dovich(1970)]{zeldovich2}
Ya. B. Zel'dovich, Astron. Astrophys. 5, 84 (1970)

\bibitem[Hawking(1974)]{hawking2}
S. W. Hawking, Nature 248, 30 (1974)

\bibitem[MacGibbon \& Carr(1991)]{macgibbon3}
J. H. MacGibbon, and B. J. Carr, Ap. J. 371, 447 (1991)

\bibitem[Halzen et al.(1991)]{halzen}
F. Halzen, E. Zas, J. H. MacGibbon, and T. C. Weekes, Nature 353, 807 (1991)

\bibitem[Page \& Hawking(1976)]{page3}
D. N. Page, and S. W. Hawking, Ap. J. 206, 1 (1976)

\bibitem[Okele \& Rees(1980)]{okele}
P. N. Okele, and M. J. Rees, Astron. Astrophys. 81, 263 (1980)

\bibitem[Cline, Matthey \& Otwinowski(2003)]{cline}
D. B. Cline, C. Matthey, and S. Otwinowski, Astropart. Phys. 18, 531 (2003)

\bibitem[Green(2001)]{green3}
A. M. Green, Phys. Rev. D 65, 027301 (2001)

\bibitem[Barrow et al.(1991)]{barrow2}
J. D. Barrow, E. J. Copeland, E. W. Kolb, and A. R. Liddle, Phys. Rev. D 43, 984 (1991)

\bibitem[Gibilisco(1996)]{gibilisco}
M. Gibilisco, Int. J. Mod. Phys. A 11, 5541 (1996)

\bibitem[Stojkovic \& Freese(2004)]{stojkovic}
D. Stojkovic, and K. Freese, Phys. Lett. B 606, 251(2005)

\bibitem[Stojkovic, Freese \& Starkman(2005)]{stojkovic2}
D. Stojkovic, K. Freese, and G. Starkman, Phys. Rev. D 72, 045012 (2005)

\bibitem[Lindley(1980)]{lindley1}
D. Lindley, Mon. Not. R. Astron. Soc. 193, 593 (1980)

\bibitem[Kaiser(1984)]{kaiser}
N. Kaiser, Ap. J. Lett. 284, L9 (1984)

\bibitem[Jensen \& Szalay(1991)]{jensen}
L. G. Jensen, and A. S. Szalay, Ap. J. Lett. 305, L5 (1986)

\bibitem[Politzer \& Wise(1984)]{PolWis}
H. D. Politzer, and M. B. Wise, Ap. J. 285, L1 (1984)


\bibitem[Sigurdsson \& Hernquist(1993)]{sigurdsson}
S. Sigurdsson, and L. Hernquist, Nature 364, 423 (1993)

\bibitem[Begelman, Blandford \& Rees(1980)]{begelman}
M. C. Begelman, R. D. Blandford, and M. J. Rees, Nature 287, 307 (1980)

\bibitem[Saslaw, Valtonen \& Aarseth(1974)]{saslaw}
W. C. Saslaw, M. J. Valtonen, and S. J. Aarseth, Ap. J. 190, 253 (1974)

\bibitem[Carr \& Lidsey(1993)]{carr10}
B. J. Carr, and J. E. Lidsey, Phys. Rev. D 48, 543 (1993)

\bibitem[Carr, Gilbert \& Lidsey(1994)]{carr2}
B. J. Carr, J. H. Gilbert, and J. E. Lidsey, Phys. Rev. D 50, 4853 (1994)

\bibitem[Kim \& Lee(1996)]{kim1}
H. I. Kim, and C. H. Lee, Phys. Rev. D 54, 6001 (1996)

\bibitem[Green \& Liddle(1997)]{GreLid}
A. M. Green, and A. R. Liddle, Phys. Rev. D 56, 6166 (1997)

\bibitem[Green, Liddle \& Riotto(1997)]{green1}
A. M. Green, A. R. Liddle, and A. Riotto, Phys. Rev. D 56, 7559 (1997)

\bibitem[Kim, Lee \& MacGibbon(1999)]{kim2}
H. I. Kim, C. H. Lee, and J. H. MacGibbon, Phys. Rev. D 59, 063004 (1999)

\bibitem[Kribs, Leibovich, \& Rothstein(1999)]{kribs}
G. D. Kribs, A. K. Leibovich, and I. Z. Rothstein, Phys. Rev. D 60, 103510

\bibitem[Bugaev \& Konishchev(2002)]{bugaev}
E. V. Bugaev, and K. V. Konishchev, Phys. Rev. D 66, 084004 (2002)

\bibitem[He \& Fang(2002)]{he}
P. He, and L. Fang, Ap. J. Lett. 568, 1 (2002)

\bibitem[Chisholm(2006)]{chisholm}
J. R. Chisholm, in preparation (2006)


\bibitem[Weinberg(2004)]{weinberg2004}
S. Weinberg, Phys. Rev. D 70, 043541 (2004)

\bibitem[Coles \& Jones(1991)]{coles}
P. Coles, and B. Jones, Mon. Not. R. Astron. Soc. 248, 1 (1991)

\bibitem[Carr \& Silk(1983)]{carrsilk}
B. J. Carr, and J. Silk, Ap. J. 268, 1 (1983)

\bibitem[Bean \& Magueijo(2002)]{bean}
R. Bean, and J. Magueijo, Phys. Rev. D 66, 063505 (2002)

\bibitem[D\"{u}chting(2004)]{duchting}
N. D\"{u}chting, Phys. Rev. D 70, 064015 (2004)

\bibitem[Cust\'{o}dio \& Horvath(2005)]{custodio1}
P. S. Cust\'{o}dio, and J. E. Horvath, Int. J. Mod. Phys. D 14, 257 (2005)

\bibitem[Peiris et al.(2003)]{peiris}
H. V. Peiris et al., Ap. J. Supp. 148, 213 (2003)

\end{thebibliography}
\end{document}